\documentclass{article}
\usepackage[utf8]{inputenc}
\usepackage{charter} 
\usepackage{geometry}
\usepackage{authblk}
\usepackage{siunitx}
\usepackage{amsfonts}
\usepackage{hyperref}
\usepackage{graphicx}
\usepackage{xcolor}
\graphicspath{ {./figures/} }
\usepackage[style=nature]{biblatex}
\DeclareUnicodeCharacter{2212}{*************************************}

\renewcommand{\vec}[1]{\mathbf{#1}}

\title{Quantifying Disorder One Atom at a Time Using an Interpretable Graph Neural Network Paradigm}

\author[1]{James Chapman\thanks{Corresponding Author, chapman37@llnl.gov}}
\author[2]{Tim Hsu\thanks{Corresponding Author, hsu16@llnl.gov}}
\author[2]{Xiao Chen}
\author[1]{Tae Wook Heo}
\author[1]{Brandon C. Wood\thanks{Corresponding Author, wood37@llnl.gov}}

\affil[1]{Materials Science Division, Lawrence Livermore National Laboratory, Livermore, CA, USA}
\affil[2]{Center for Applied Scientific Computing, Lawrence Livermore National Laboratory, Livermore, CA, USA}

\begin{document}

\maketitle

\begin{abstract}

Quantifying the level of atomic disorder  within materials is critical to understanding how evolving local structural environments dictate performance and durability. Here, we leverage graph neural networks to define a physically interpretable metric for local disorder. This metric encodes the diversity of the local atomic configurations as a continuous spectrum between the solid and liquid phases, quantified against a distribution of thermal perturbations. We apply this novel methodology to three prototypical examples with varying levels of disorder: (1) solid-liquid interfaces, (2) polycrystalline microstructures, and (3) grain boundaries. Using elemental aluminum as a case study, we show how our paradigm can track the spatio-temporal evolution of interfaces, incorporating a mathematically defined description of the spatial boundary between order and disorder. We further show how to extract physics-preserved gradients from our continuous disorder fields, which may be used to understand and predict materials performance and failure. Overall, our framework provides an intuitive and generalizable pathway to quantify the relationship between complex local atomic structure and coarse-grained materials phenomena.








\end{abstract}
\section{Introduction}

\begin{figure}
    \centering
    \includegraphics[width=0.85\textwidth]{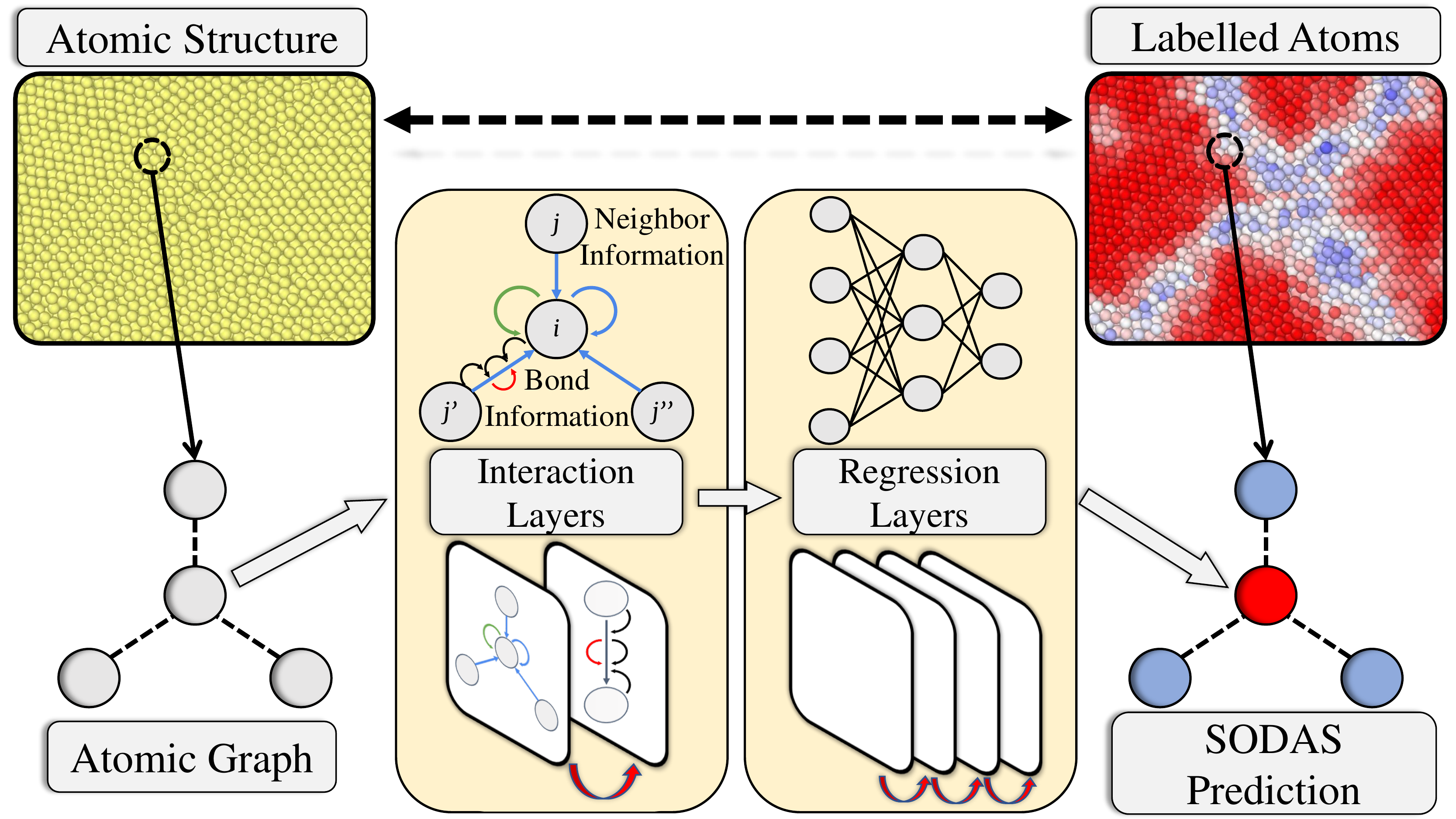}
    \caption{General workflow for calculating SODAS values. Atomic structures are converted into graph representations, which explicitly encode all necessary geometric information. These atomic graphs are then fed into a graph neural network, which has been trained to distinguish between the unique local geometries in different material phases. The graph neural network then gives each atomic environment a SODAS value, which classifies where in the phase space, between phases, that local structure is most likely to occur.}
    \label{fig:workflow}
\end{figure}

Understanding how a material's structure affects its properties is one of the most fundamental principles in materials science. At the center of this paradigm is the fact that macroscopic materials behavior begins at the atomic scale, with local atomic arrangements ultimately coming together to form structural features observed at larger length scales \cite{Fish2021,Miller_2009}. Characterizing the nature and propagation of these local environments is therefore vital to understanding macroscale structure-property relationships and their evolution \cite{KARMA201625}. Complicating this endeavor is the fact that the long-range features often depend on structurally disordered atomic environments, which tend to dictate materials functionality \cite{GUO2016624}. For instance, transport, chemical reactivity, and phase nucleation are all profoundly affected by the presence of interfaces, interphases, and grain boundaries \cite{doi:10.1126/science.1078616,https://doi.org/10.1002/adma.200800440,doi:10.1021/la052551e,Heo2021,HEO201468,HEO2019262}. These processes in turn are intricately connected to performance-durability trade-offs in both functional \cite{GITTLEMAN201981} and structural \cite{https://doi.org/10.1002/adma.201805876} materials.  Examples include temperature-dependent microstructure evolution \cite{doi:10.1063/1.2423084,LIU2021110733}, hotspot formation \cite{SIMON2010106,doi:10.1021/j100111a031}, and the nucleation and growth of new material phases \cite{BUDEVSKI20002559,doi:10.1021/cr2001756}. 

However, quantifying local atomic disorder in a physically motivated way in practice is extraordinarily difficult \cite{https://doi.org/10.1002/anie.201902625}. Although a number of methods have been proposed to characterize local atomic environments, these methods are often not optimized to magnify the subtle differences present in disordered environments. Existing methods can typically be grouped into three general classes, each of which carries distinct trade-offs: (1) \textit{semi-empirical structure factors} such as Adaptive Common Neighbor Analysis (CNA) \cite{Stukowski_2012}, Steinhardt order parameters (SOP) \cite{doi:10.1080/01418618108235816}, Ackland-Jones order parameters (AJ) \cite{PhysRevB.73.054104}, atomic excess volume \cite{Mahmood2022}, atomic "Smoothness" metrics (ASM) \cite{doi:10.1073/pnas.1807176115}, and the local atomic environment metric (LAE) \cite{Rosenbrock2017}; (2) \textit{parameterized symmetry functions} such as the Smooth Overlap of Atomic Positions (SOAP) \cite{C6CP00415F}, Behler-Parinnello functions (BP) \cite{doi:10.1063/1.3553717}, Moment Tensor Representations (MTR) \cite{doi:10.1137/15M1054183}; and the Adaptive Generalizable Neighborhood Informed functions (AGNI) \cite{CHAPMAN2020109483}; and (3) \textit{unsupervised machine learning methods} which include graph-based \cite{Park2021} and image-based \cite{Chan2020} representations. 

In general, it is highly desirable to develop a methodology that is by construction specifically designed to distinguish, quantify, and physically interpret regions with varying degrees of atomic disorder. Such a capability would enable more accurate predictions of how disordered atomic environments translate to higher-level features and functionality. For instance, mapping between discretized atomistic models and continuous field representations, such as phase-field \cite{Steinbach_2009,doi:10.1146/annurev.matsci.32.112001.132041,doi:10.1146/annurev.matsci.32.101901.155803} and finite-element \cite{https://doi.org/10.1002/ar.a.20169} models, forces the use of ill-defined and arbitrary approximations \cite{KAVOUSI2022126461}, particularly when disorder is present. Moreover, continuous field representations propagate via local gradients \cite{doi:10.1080/00018730701822522}, the evaluation of which amplifies inaccuracies associated with disordered regions. Addressing these shortcomings is therefore a critical priority.

To this end we introduce a physics-aware workflow composed of two stages, which can be seen in Fig. \ref{fig:workflow}. First, we use graph neural networks (GNN) to explicitly encode local atomic structural information. Next, we apply this encoding to map the local atomic structure to a novel order parameter that characterizes local disorder. This order parameter, henceforth referred to as the Structural Orderness Degree for Atomic Structures (SODAS), $\lambda_{i}$, quantifies an atom's local structure in terms of the ``closeness'' to likely environments encountered between two limiting cases: a perfect crystal ($\lambda_{i} =$ 1) and a melt ($\lambda_{i} =$  0). Our approach offers three distinct advantages: (1) the graph representation accurately encodes the topology of the connected network of atoms; (2) SODAS is well-defined and bounded, facilitating universal and intuitive applicability across all atomic structures of a material; and (3) the compact GNN encoding allows for physical interpretability. The power of this workflow is demonstrated by application to several examples of disordered aluminum systems including solid-liquid interfaces, polycrystalline microstructures, and grain boundaries.

\section{Results}

\subsection{Definition of SODAS}

In principle, the level of configurational disorder can be mapped onto an equivalent level of thermal disorder in a finite-temperature ensemble. To this end, we can introduce a fictitious temperature ($T'$) that mathematically represents this configurational disorder. In practice, $T'$ can be parameterized for a given system using explicit MD simulations as discussed in the Methods section. To physically bound $T'$, we introduce $T_{d}$ as the limit of full disorder (nominally the melting temperature). The value of $T'$ is then confined to the range between 0 and $T_{d}$. We next define $\gamma$ as a thermodynamic order parameter:

\begin{equation}
    \gamma (T'; T_d, s) =  \mathcal{N} \frac{1}{1 + \exp(-(T_d / T')^{s})}
\end{equation}


where $\mathcal{N}$ normalizes $\gamma$ between 0 (absolute disorder) and 1 (absolute order), and $s$ is an empirical scaling metric that determines where to begin the decay of $\gamma$ from ordered to disordered. The introduction of $s$ makes the definition of $\gamma$ universal, as one can simply tailor its value for any unique material system. A plot of the relationship between $\gamma$ and $T'$ can be found in the supplemental information. 


While $\gamma$ describes the level of disorder of a macroscopic, homogeneously disordered system, we are primarily interested in local atomic disorder within a \textit{heterogeneous} system. To establish this connection, we map the likelihood of finding a given local atomic environment within an ensemble of configurations, to a local order parameter (SODAS), $\lambda(n)$, where $n$ indexes an atom. In practice, due to ergodic constraints, we assume that this ensemble can be sampled discretely from MD simulations. Note that this is analogous to determining an atomically resolved configurational entropy density. We represent this mapping as:

\begin{equation}
     f(\{\gamma\}_{i}) \mapsto \lambda(n)
\end{equation}

where $\{\gamma\}_{i} =  \{\gamma_{1},\gamma_{2},...,\gamma_{k}\}$ represents the set of $\gamma$ values associated with a given local atomic structural motif $i$ across a discrete set of $k$ ensembles, and $f$ is a function that maps $\{\gamma\}_{i}$ to $\lambda(n)$. While the function $f$ is unknown, it can be approximated. In this work we use a graph neural network scheme to facilitate this approximation, while retaining physical interpretability. Fig. \ref{fig:workflow} outlines the key steps in this process and is discussed in further detail in the methods section.



\subsection{Validation of SODAS}

\begin{figure}
    \centering
    \includegraphics[trim={0 6cm 0 0},width=0.8\textwidth]{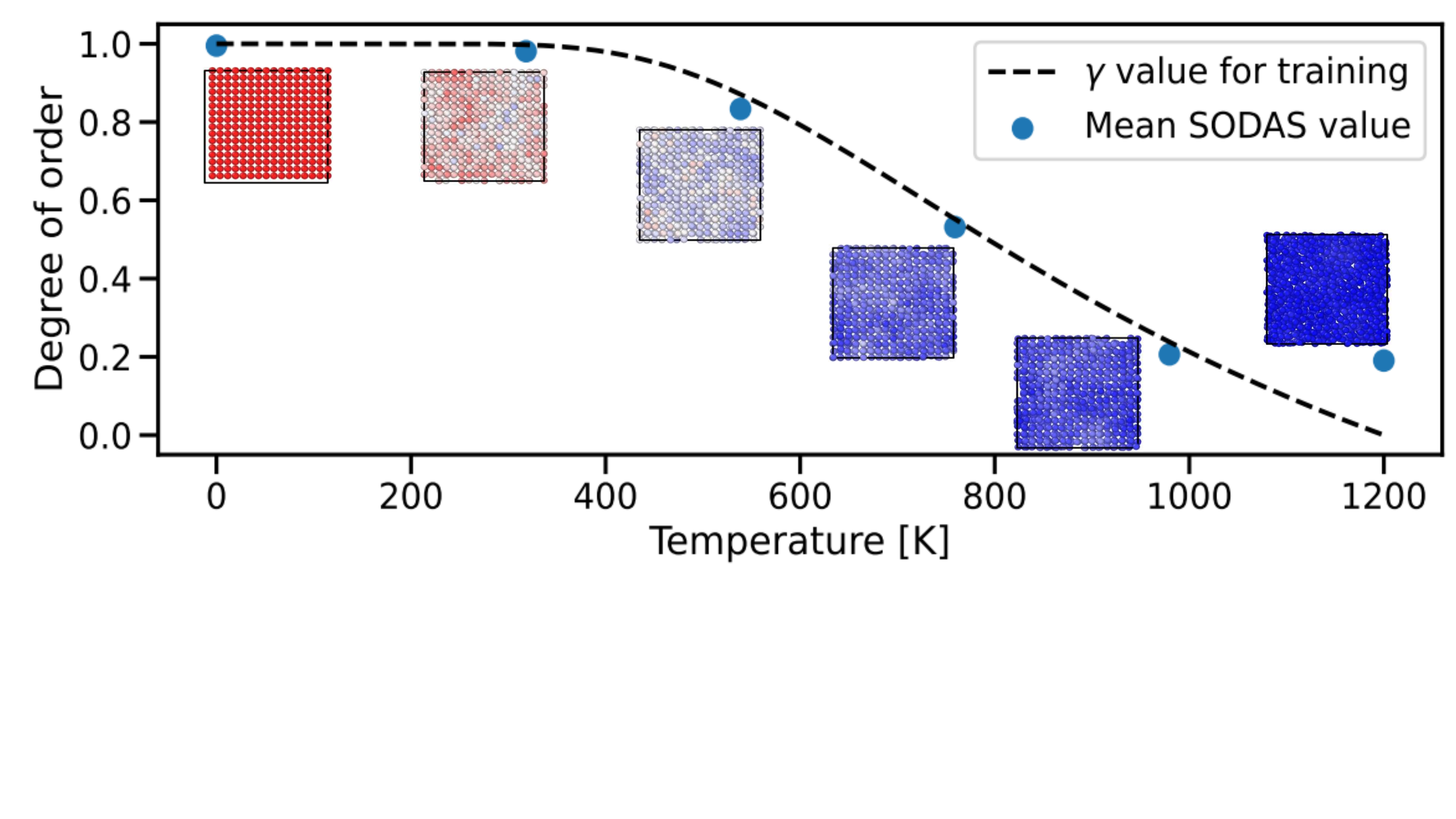}
    \caption{SODAS calculations on bulk structures taken during a superheating MD simulation. Values along the y-axis represent the average SODAS value for each shown structure, whose atoms are colored according to each atom's SODAS value. The dashed line indicates the theoretical values of $\gamma$ while the plotted SODAS values represent the accuracy of the GNN mapping.}
    \label{fig:val_data}
\end{figure}

We first validate SODAS by applying our procedure to molecular dynamics simulations of pristine bulk FCC Al at temperatures up to 1200K (melt). Fig. \ref{fig:val_data} provides a visual depiction of the $\lambda(n)$ mapping with respect to the theoretical prediction of $\gamma$. Here we see that all atoms in the structure at 0K is uniformly predicted to have $\lambda = $ 1, which is indicative of the perfect crystal. In contrast, at 1200K, all atoms indicate $\lambda$ to be close to 0 due to the structure existing as a melt. On average, structures between these limits yield intermediate values of $\lambda$, as expected.

At the same time, detailed visualization of intermediate-temperature configurations reveals a spectrum of atomic environments covering a range of $\lambda(n)$ in lieu of homogeneously distributed disorder. For example, the second structure in Fig. \ref{fig:val_data}, which represents a structure at roughly 200K, has $\lambda$ values ranging from 0.5 to 0.95. Accordingly, as described previously, similar atomic environments exist at a range of temperatures but with different degrees of expression according to the average overall level of disorder. Intuitively this makes sense, as the goal of the SODAS metric is to judge the likelihood of an atomic environment existing at an arbitrary point along the thermodynamic spectrum between fully ordered and fully disordered variants. If a unique atomic environment occurs at multiple temperatures, one would expect its $\lambda$ to be a weighted combination of the individual occurrences of the environment along the temperature spectrum. 

Importantly, Fig. \ref{fig:val_data} also compares the GNN-learned SODAS values with respect to the theoretical values for $\gamma$. Here, we can see excellent agreement between the GNN mapping and $\gamma$ up to around $T = $1000K. This indicates that our MD simulations are sufficient to capture the configurational entropy present within the material at these temperatures. However, deviations exist above $T = $1000K, with a nearly identical average SODAS value predicted within this temperature range. This can be explained by our choice of $T_d =$ 1200K (slightly above the EAM melting temperature $\sim$1050K) to improve sampling in the high-temperature limit; further details can be found in the supplemental information.



\subsection{Boundary Identification in Solid-Liquid Interfaces}

While crystalline interfaces are easy to identify, disordered interface boundaries are intrinsically more complex, making classification of the transition into the boundary region ill-defined. Bond-angle methods such as CNA and AJ, and more complete methods such as SOAP, often fail to distinguish between perturbed crystalline and disordered atomic environments. These difficulties ultimately make defining interface boundaries challenging. In contrast, the SODAS formalism accomplishes this by providing a \textit{continuous} metric that allows for a physically justifiable and mathematically rigorous definition of the interface boundary transition.

\begin{figure}
    \centering
    \includegraphics[trim={2cm 4cm 2cm 0},width=0.7\textwidth]{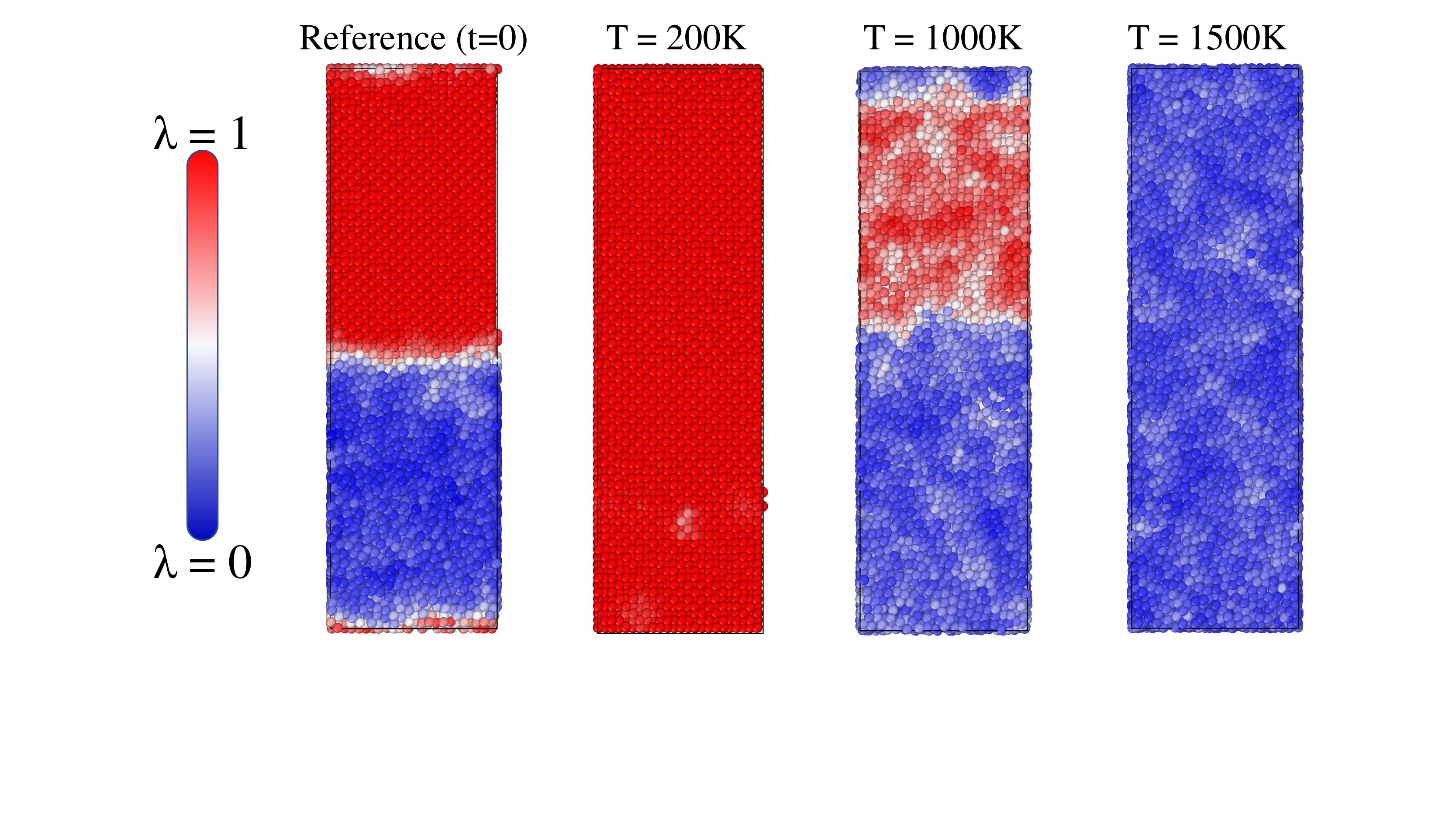}
    \caption{Snapshots taken from MD 2-phase simulations. The reference structure is taken at the moment the 2 phases are allowed to coexist ($t = 0$). The 200K, 1000K, and 1500K snapshots are taken at the end of each MD simulation. The atoms in each system are color-coded based on their SODAS values, as reference din their colorbar.}
    \label{fig:2phase}
\end{figure}

To this end, we have performed two-phase crystal/liquid CMD simulations at several temperatures (200K, 1000K, and 1500K), to observe how SODAS classifies the unique structural environments present in each scenario. Further details regarding the simulation setup can be found in the Methods section. Fig. \ref{fig:2phase} shows the SODAS characterization of the atomic environments present for the three temperatures. For the CMD simulation at 200K SODAS correctly identifies the collapse of the interface boundary. It also locates a small cluster of local disorder, indicating the presence of a point defect-like region, as a vestige of solidification of the liquid region. Analogously, at 1500K SODAS correctly identifies the existence of a single liquid phase and the loss of the boundary region. SODAS does however, predict varying levels of disorder as well as the presence of pockets within the liquid that show moderate-to-low disorder.


On the other hand, at 1000K SODAS correctly captures the interface between the two phases, as well as moderate perturbations relative to the reference structure. More importantly, we also see that the interface between the two phases is not a straight line, instead reflecting the more jagged, realistic topology of the boundary. SODAS further captures the volumetric nature of the entire interface boundary region's complex morphology, which features spatio-temporally varying 3D shape and thickness. 

\subsection{Autonomous Microstructural Feature Extraction}

The ability to define boundary transitions also enables the identification of larger-scale microstructural features. To demonstrate this capability, we showcase how a combination of SODAS and a graph characterization algorithm can autonomously quantify complex polycrystalline microstructures as they evolve through both time and temperature domains. As described in the Methods section, we performed three MD simulations of a 1.6 million atom FCC aluminum system containing 5, 50, and 250 initial grains, respectively. 

\begin{figure}
    \centering
    \includegraphics[width=0.8\textwidth]{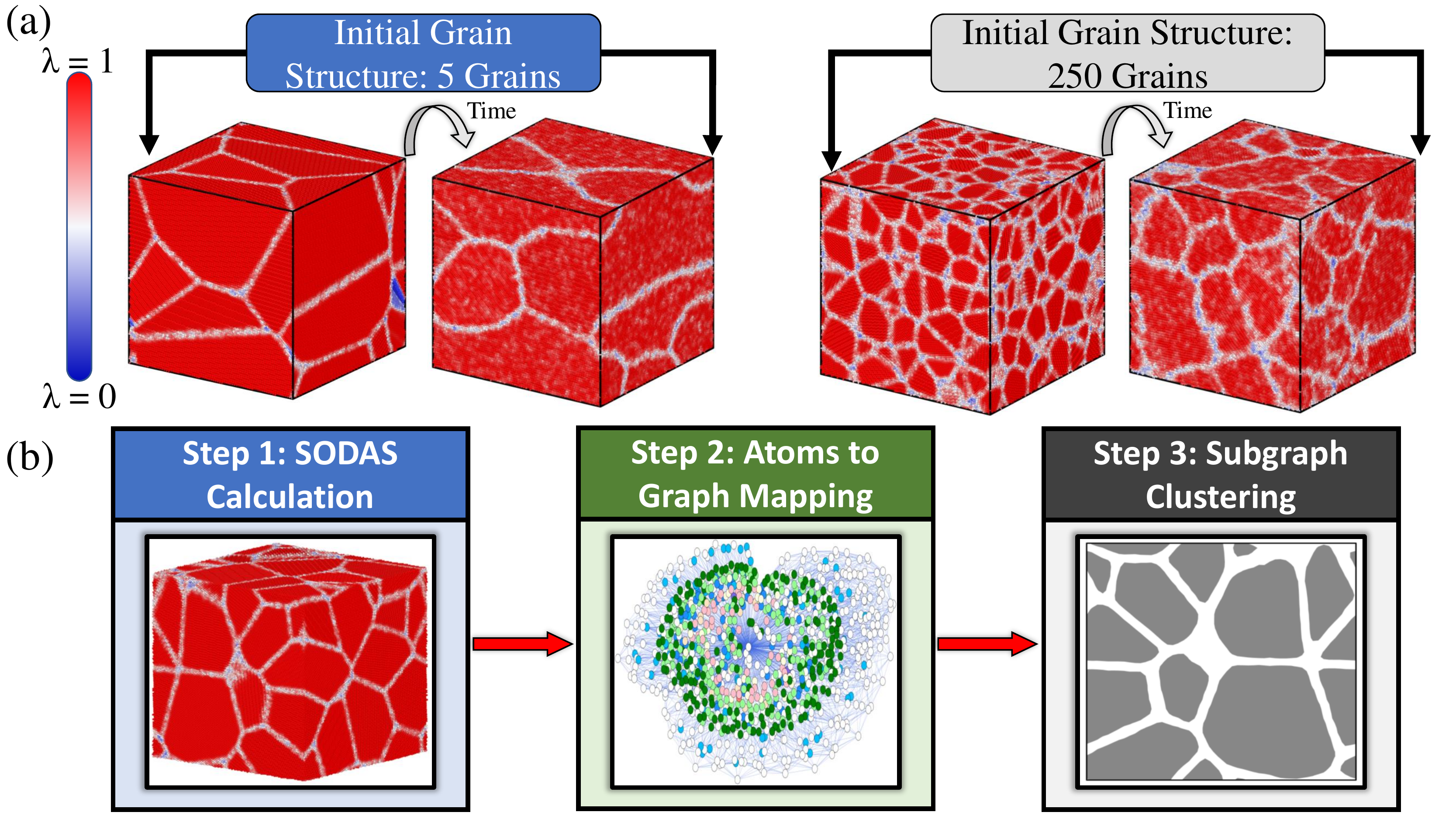}
    \caption{(a) SODAS predictions on the initial and final configurations of the polycrystalline MD simulations. Each column block represents the initial grain structure present in the system, with the right-side representing the 5 initial grains, and the left side indicative of the 250 initial grains. Each row, color-coded in green and yellow, represents the temperature of the MD simulation, with green representing 200K and yellow being 600K. (b) Workflow of the unsupervised graph-based grain detection algorithm, visualized using simulated polycrystalline Al. (Left) SODAS values for each atomic environment present in the system. (Middle) Atoms-to-graph mapping (after SODAS thresholding), where node colors represent the connectivity of a given atom. (Right) Autonomous grain detection using recursive subgraph clustering.}
    \label{fig:micro-view}
\end{figure}

Figure \ref{fig:micro-view} (a) provides a visual depiction of how one can identify the changes in microstructure as a function of time, with examples of polycrystalline Al systems containing 5 and 250 initial grains.
In Fig. \ref{fig:micro-view} (b), we provide a visual workflow that shows how autonomous microstructural feature extraction is performed using three steps. First, SODAS calculations are performed in which each atom is assigned a SODAS value as described earlier. Second, an atomic graph is constructed, where all atoms below a SODAS threshold are removed and the remaining atoms are converted to an atomic graph \cite{chapman2021sgop} in order to remove grain boundary atoms and leave grain atoms. Third, grain characterization is performed through subgraph clustering, where an exhaustive, recursive subgraph search is performed on the atomic graph to identify the unique grains. Further details can be found in the Methods. Importantly, this procedure leverages a unique feature of SODAS---namely, the ability to provide a physically-motivated and mathematically well-defined threshold for identifying which atoms belong to the grain and which belong to the grain boundary. This process is extremely challenging using existing methods due to their lack of a continuously and bounded metric.



\begin{figure}
    \centering
    \includegraphics[width=1.0\textwidth]{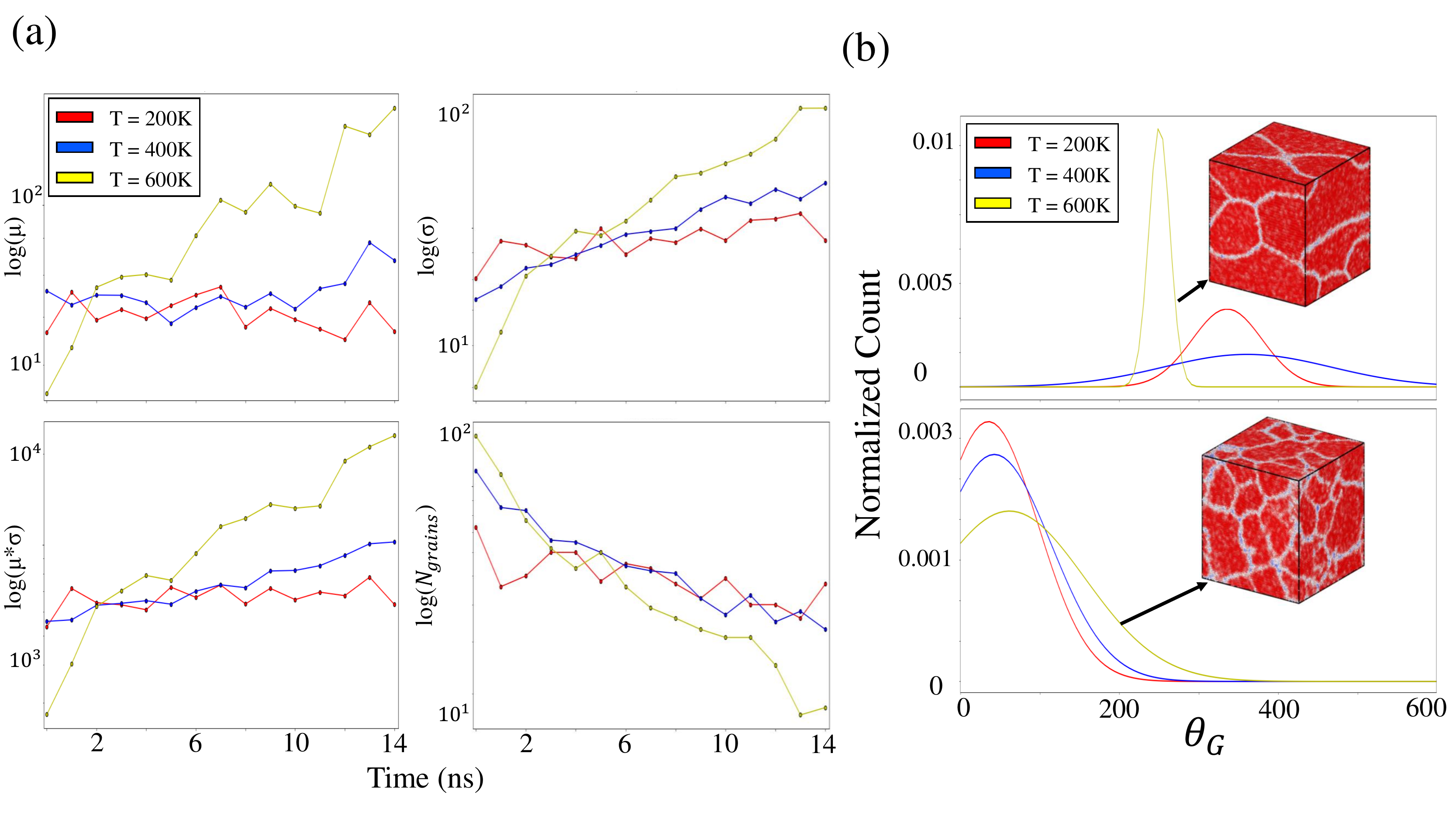}
    \caption{Quantitative metrics characterizing microstructural changes over time. In each case the colors correspond to the temperature of the MD simulation. (a) Changes in the graph order parameter distribution's (top, left) mean, (top, right) standard deviation, (bottom, left) multiplication of the mean and  standard deviation, and (bottom, right) number of grains, all as a function of time, starting from the 250 grain structure. (b) SGOP microstructure characterization performed the SODAS predicted values for the case of (top) 5 initial grains and (bottom) 250 initial grains. Histograms are determined using the resulting SGOP values, calculated on the final structure from each MD simulation (the last point in each subplot in (a)). Inserted images, and corresponding arrow, show the final structure obtained from the 600K MD simulations. All histograms are normalized and show each x-value's probability.}
    \label{fig:micro-char}
\end{figure}

Fig. \ref{fig:micro-char} provides a quantitative understanding of how the grain structure changes as both a function of the initial grain morphology, as well as the temperature. To quantify this relationship, we invoke our recently developed graph order parameter (SGOP) \cite{chapman2021sgop}. Here, we emphasize that SGOP provides an approximate measure to a grain's atomic-level connectivity, which in turn encodes the ratio between near-boundary and interior atoms, and can therefore serve as a proxy to that grain's shape and size. As a result, the distribution of SGOP values provides a physically-intuitive and robust measure of the microstructure. This featurization is a more accurate and unique representation compared to the number of atoms in a grain, the grain radius, or the grain density.

Figure \ref{fig:micro-char} (a) provides histograms which are generated using the SGOP values calculated on the final structure from the MD simulations performed at 200K, 400K, and 600K. Here, one can observe different interplay between temperature and grain size for two examples containing 5 and 250 initial grains. For the former case (top), at 200K there is an abundance of larger grains, as little-to-no grain coarsening occurred. At 400K, one can see a broadening of the distribution, indicating the coexistence of grains both smaller and larger than those existing at 200K. The same trend continues at 600K, yielding more medium sized grains. For the latter case of 250 initial grains (bottom), the distributions are more consistent, indicating a more homogeneous spread of grain shapes and sizes. Nevertheless, temperature effects continue to broaden the distributions. 

To interpret the results of (a) we track the time-evolution of the microstructure. To do so, we use a normal distribution to fit the SGOP distributions, whose mean and standard deviation can be used as a microstructural feature vector, as shown for the 250-grain case in Figure \ref{fig:micro-char} (b). At 200K there is initially an increase in the mean, shown in Fig. \ref{fig:micro-char} (b,top-left), implying that larger, more connected grains exist throughout the structure. However, a plateau is reached within 2 ns, and no further increase in the mean can be seen. At 400K a similar trend is observed, though some further growth in the grains exists between 10 and 14 ns. At 600K one can observe a constant trend, indicating that grain growth occurs over the course of the entire simulation.

Changes in the standard deviation, seen in Fig. \ref{fig:micro-char} (b,top-right), can be attributed to grain coarsening, because as one grain grows in size, another neighboring grain must reduce in size as it is swallowed by the growing grain. This process ultimately causes a fluctuation in the spread of graph order parameter values, which is captured by the standard deviation of the fitted distribution. At 200K there is initially an increase in the standard deviation, indicating the presence of grain coarsening. However, in agreement with the mean values, a plateau is reached within 2 ns, and no further increase in the standard deviation can be seen. At 400K, one can observe a constant increase in the standard deviation, indicating that grain coarsening is occurring throughout the course of the trajectory. The slope of this line can be qualitatively linked to the rate of coarsening. At 600K the trend is similar to 400K but with a larger slope, indicating that grain coarsening at 600K is occurring more rapidly.

Figure \ref{fig:micro-char} (b,bottom-left) combines the mean and standard deviation into a single metric. Here, the product of the two terms is used to gauge both the rate of grain growth and the magnitude of the grains themselves. At 200K we see a trend that confirms our previous observation that grain growth initially occurs rapidly, but tapers off quickly. At 400K we observe a more complete picture, where slow grain growth occurs throughout the trajectory. For the case of 600K, faster growth can be seen, as well as the existence of larger grains. 

This picture can also be complemented by observing how the number of grains changes as a function of time, as seen in Fig. \ref{fig:micro-char} (b,bottom-right). We can see that at 200K, there is an initial reduction in the number of grains from 250 to approximately 75, where that number holds for the remainder of the simulation. At 400K there is a constant reduction in the number of grains from 250 roughly 30 by the end of the MD simulation. At 600K the number of grains is reduced even further by the end of the simulation, going from 250 to only 12 grains. The picture painted in Fig. \ref{fig:micro-char} (b,bottom-right) aligns well with the values in Fig. \ref{fig:micro-char} (b,bottom-left), indicating that our product metric can be used to quantify both the size, shape, and the growth rate of the grains.

\subsection{Mapping Atoms to a Continuous Field}

Lastly, since $\lambda$ is continuously valued over the discrete atoms, it can be interpolated to a continuous field. This mapping allows for its integration into continuum models. For instance, we note the similarity between such a continuous field representation and the phase order parameter used in phase field models \cite{Heo2021,Heo_2021}. We showcase this concept for the example of two grain boundary regions with varying levels of interfacial complexity. Nevertheless, we note that this method can be used for other classes of crystalline interfaces, such as symmetric tilt and twin boundaries, and edge/screw dislocations.

\begin{figure}
    \centering
    \includegraphics[trim={0 1cm 0 0}, width=0.9\textwidth]{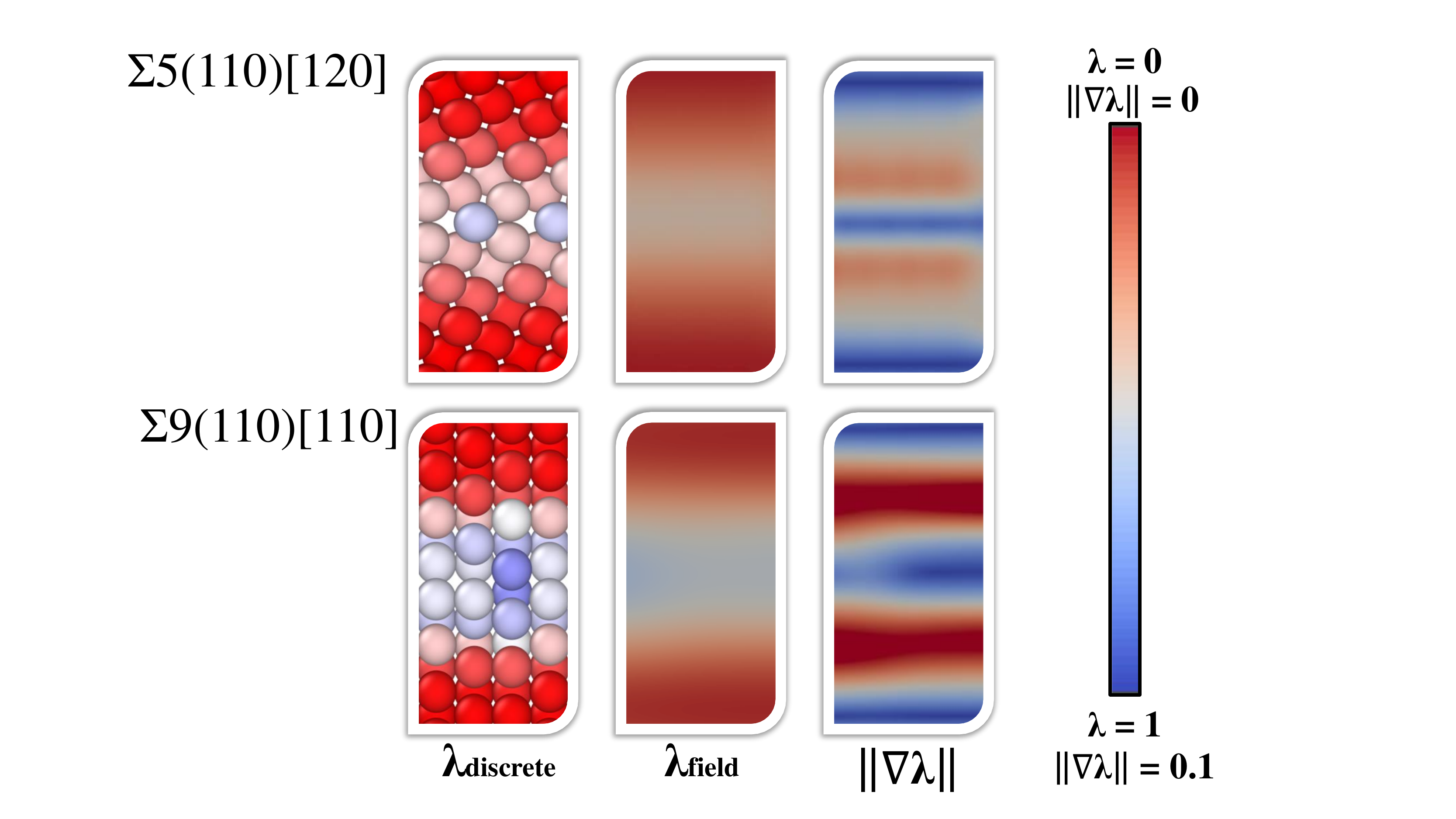}
    \caption{Continuous fields (and its gradient norm) of the originally discrete, per-particle SODAS value $\lambda$. The discrete-to-continuum conversion is done by interpolating the discrete $\lambda$ onto a uniform, fine grid. The gradient information can then be computed over the uniform grid. }
    \label{fig:continuous-example}
\end{figure}

From Fig. ~\ref{fig:continuous-example}, one can see the intuitive nature of SODAS, cleanly characterizing the grain regions with a $\lambda$ close to 1, smoothly transitioning to higher degrees of disorder present near the boundary. For boundaries that show higher degrees of crystallinity, such as those in $\sum5(110)[120]$, the disorder present at the interface is minimal, as is expected, though is still clearly present. Likewise, for more disordered boundaries, such as those in $\sum9(110)[110]$, a greater degree of disorder is detected within the interface region. As in the previous section, these characterizations exemplify the ability of SODAS to intuitively determine where the grain begins and ends.

Figure~\ref{fig:continuous-example} shows the continuous fields derived from the originally discrete, per-particle SODAS value $\lambda$. Additionally, the gradient norm $||\nabla \lambda||$ was calculated and visualized. This discrete-to-continuum conversion was done by interpolating the discrete $\lambda$ values onto a uniform grid using PyVista \cite{sullivan2019pyvista}. When calculating the gradient of this field we observe areas of the structure where there are sharp changes in the SODAS values. Notably, the gradient is maximized not at the center of the grain boundary, but rather at the transition to the boundary region, because these are locations within the structure where there is an abrupt change in the level of disorder present. 

The sensitivity of this detection can be seen in Fig.~\ref{fig:continuous-example}, where the gradient of the scalar field predicts two regions where there is an abrupt change in the SODAS values. As we move from the crystalline regions towards the interface normal to the boundary region, we first encounter a crystal-to-boundary region, followed by the boundary itself, and finally a boundary-to-crystal region as we move away from the interface. Therefore, the gradient predictions in Fig.~\ref{fig:continuous-example} highlight the fact that a degree of homogeneity can exist in both the ordered interior of the grain as well as the disordered interior of the grain boundary.



\section{Discussion and Conclusion}

In summary, characterizing the nature of local atomic disorder is critical and necessary to understand how structure-property relationships evolve. SODAS is a new mathematical framework in which local atomic environments are transformed into graph representations, encoded via a graph neural network paradigm, and finally mapped onto a local order parameter. This order parameter, $\lambda$, is a physically intuitive, continuous, and mathematically bounded scalar which represents the level of disorder present within an atomic environment, and is analogous to an atomically resolved configurational entropy density. In addition to the examples shown throughout this work, these advantages allow for the universal quantification of a multitude of complex and heterogeneous materials properties and phenomena.

We also envision our proposed methodology as a novel tool for multiscale model integration. In particular, SODAS provides an atomistically derived, physically motivated continuous scalar field representation for phase field and continuum models. This mapping can be likewise leveraged to output field quantities such as phase order, grain distribution, concentration, stress/strain, and so on. Such an approach offers a new perspective and valuable technique for bridging scales in multiscale models, both between atomistic and microscale descriptions, as well as between discrete and continuous representations. We further emphasize that although this work focuses on single-element systems, our method is generally applicable to multi-component systems and their corresponding microstructural features.

The advantages of SODAS also become clear for extraction of physical properties that relate to materials performance or degradation. For instance, we showed that by interpolating the discrete representation to a continuum representation, we can analyze or differentiate $\lambda$ to deduce spatially resolved changes in structural homogeneity. In practice, these structural changes often map to changes in key response properties, including diffusivity, dielectric response, electrical conductivity, and elastic compliance.~\cite{heo2021microstructural} In cases where such properties can be computed locally or measured using local probes, SODAS offers a way to extract analytical relationships between structure and function. Moreover, sharp gradients from abrupt changes in response functions can concentrate electrical, chemical, or mechanical potential, creating ``hotspots'' that can initiate key electrochemomechanical failure modes. We therefore propose that gradients in the continuous representation of $\lambda$ may provide a unique way to identify such ``hotspots'', with direct connection to early prediction of propensity for deleterious outcomes such as fracture, corrosion, and thermal runaway.




\section{Methods}

\subsection{Training Data Preparation}

Classical molecular dynamics (CMD), using the LAMMPS software package \cite{PLIMPTON19951}, was used to generate training data for the GNN model. Starting from bulk FCC aluminum (containing 1024 atoms), CMD was performed in the NVT ensemble for 1 nanosecond using the Zhou et al EAM potential \cite{ZHOU20014005}, increasing the temperature linearly with respect to time. The initial and final tempeartures for the CMD simulation was 100 and 1200, respectively. 1000 equidistant snapshots were then taken along the trajectory and labelled according to their respective temperatures. Further information regarding the training data preparation can be found in the supplemental information.

During training, all atoms at a given temperature are assigned the same value of $\gamma$. In this way, the value of $\gamma$ is not directly connected to an atom's local structural geometry. However, at each temperature local atomic geometries bounce around an equilibrium point, which in this case can be thought of as the 0K structure. As the temperature increases the magnitude of the displacement from this structure increases, though some atomic neighborhoods may resemble low temperature structural motifs, even at higher overall temperatures.

\subsection{Graph Neural Network Implementation}
    
    \subsubsection{Conversion to Graph}
    Prior to GNN operation, we converted the atomic systems into graphs by a simple cutoff radius-based neighbor list search (implemented using Atomic Simulation Environment \cite{ase-paper}), with the cutoff $R_c = 3.5$ \AA. Each node of the converted graph corresponds to the atom type $z$, and each edge the bond distance $d$.

    \subsubsection{GNN Operation}
    The GNN model used in this work consists of three components: the initial embedding, the atom-bond interactions, and the final output layers (Fig.~\ref{fig:workflow}).
    
    
    In the initial embedding, each atom type $z$ is transformed into a feature vector by an \verb|Embedding| layer (PyTorch \cite{NEURIPS2019_9015}). Each bond distance $d$ is expanded into a $D$-dimensional feature vector by the Radial Bessel basis functions (RBF) \cite{klicpera2020directional}
    
    \begin{equation} \label{eq:rbf}
        \text{RBF}_n(d) =
        \sqrt{\frac{2}{R_c}} \frac{\sin(\frac{n \pi}{R_c} d)}{d},
    \end{equation}
    where $n \in [1..D]$ and $R_c$ is the cutoff value. Both atom and bond feature vectors have the same length $D = 100$.
    
    The atom-bond interactions are also known as GNN convolution, aggregation, or message-passing. There are many variants of GNN convolution operations that can be adopted from the literature. In this work, we choose the edge-gated graph convolution \cite{bresson2017residual, dwivedi2020benchmarking}. The term \emph{atom-bond interaction} is based on the fact that the nodes and the edges exchange information during the convolution operation. Specifically, the node features $\vec{h}^{l+1}_i$ of node $i$ at the $(l+1)$th layer is updated as
    \begin{equation} \label{eq:node-update}
        \vec{h}^{l+1}_i =
        \vec{h}^l_i + 
        \mathrm{SiLU} \left(
            \mathrm{LayerNorm} \left(
                \vec{W}^l_s \vec{h}^l_i + 
                \sum_{j \in \mathcal{N}(i)} \hat{\vec{e}}^l_{ij} \odot \vec{W}^l_d \vec{h}^l_j
            \right)
        \right),
    \end{equation}
    where SiLU is the Sigmoid Linear Unit activation function \cite{ELFWING20183}; LayerNorm is the Layer Normalization operation \cite{NEURIPS2019_2f4fe03d}; $\vec{W}_s$ and $\vec{W}_d$ are weight matrices; the index $j$ denotes the neighbor node of node $i$; $\hat{\vec{e}}_{ij}$ is the edge gate vector for the edge from node $i$ to node $j$; and $\odot$ denotes element-wise multiplication. The edge gate $\hat{\vec{e}}^{l}_{ij}$ at the $l$th layer is defined as
    \begin{equation} \label{eq:edge-gate}
        \hat{\vec{e}}^l_{ij} = 
        \frac
        {\sigma(\vec{e}^l_{ij})}
        {\sum_{j' \in \mathcal{N}(i)} \sigma(\vec{e}^l_{ij'}) + \epsilon},
    \end{equation}
    where $\sigma$ is the sigmoid function, $\vec{e}^l_{ij}$ is the original edge feature, and $\epsilon$ is a small constant for numerical stability. The edge features $\vec{e}^l_{ij}$ is updated by
    \begin{equation} \label{eq:edge-update}
        \vec{e}^{l+1}_{ij} =
        \vec{e}^l_{ij} + 
        \mathrm{SiLU} \left(
            \mathrm{LayerNorm} \left(
                \vec{W}^l_g \vec{z}^l_{ij}
            \right)
        \right),
    \end{equation}
    where $\vec{W}_g$ is a weight matrix, and $\vec{z}_{ij}$ is the concatenated vector from the node features $\vec{h}_i$, $\vec{h}_j$, and the edge features $\vec{e}_{ij}$:
    \begin{equation}
        \vec{z}_{ij} = \vec{h}_i \oplus \vec{h}_j \oplus \vec{e}_{ij}.
    \end{equation}
    
    Lastly, in the final output layers, each node feature is eventually transformed into a scalar output $y$ ranging from 0 to 1. Effectively, the GNN predicts the SODAS metric for every atom.
    
    Further details regarding model training are described in Supporting Information.

\subsection{Atomistic Simulation Details}

\subsubsection{2 Phase Simulations}

2 phase CMD simulations were performed on a orthogonal block of aluminum containing 23,040 atoms. All CMD simulations were performed in the NVT ensemble. Initially, half the box was superheated to 4000K to assure complete melting, while the remaining half of the box was set at 200K. This initial CMD simulation was performed for 1 nanosecond with separate NVT thermostats driving each region. The 4000K and 200K regions were then allowed to coexist using a single NVT thermostat to drive the CMD simulation. For the purpose of this analysis upon the introduction of the shared thermostat the temperature was fixed to 3 cases: one at 200K, another at 1000K, and a final case of 1500K. The purpose of these temperatures is to observe how our proposed methodology predicts the unique levels of atomic disorder present in each case. Further information regarding these simulations can be found in the Results section and in Fig. \ref{fig:2phase}.

\subsubsection{Grain Coarsening Simulations}

CMD simulations in the NVT ensemble were performed for 6 polycrystalline cases, each with a varying number of initial grains. An initial bulk aluminum system containing roughly 1.6 million atoms was used to construct 6 polycrystalline structures, using the Atomsk software package \cite{HIREL2015212}, containing 5 and 250 initial grains. CMD simulations were performed on each case at 200K, 400K, and 600K. NVT simulations were run for approximately 1.5 nanoseconds for each combination of initial structure and temperature. All CMD simulations were done using LAMMPS with the Zhou et. al EAM potential \cite{ZHOU20014005}. Further details regarding the polycrystalline structures can be found in the Results section as well as the Supplemental Information.

\subsection{Microstructure Characterization}

Microstructure characterization occurs in 4 stages: (1) calculation of SODAS for all atoms in the system, (2) thresholding of the atomic configuration, based on an atom's SODAS value and subsequent removal of all atoms below the threshold value, (3) conversion of the remaining atoms to a graph representation for the discovery of subgraphs within the graph, and (4) characterization of the grains through a graph order parameter. Fig. \ref{fig:micro-view} (b) depicts this workflow visually. While step (1) requires little-to-no input from the user, step (2) requires one to define the level of disorder that needs to be captured when defining the interface regions. This choice highlights the intuitive nature of our proposed methodology, as the threshold value defines the structural properties of the interface region itself, with a near-zero threshold indicating grain boundaries which are extremely disordered and a value close to 1 representing highly crystalline boundaries. In principle, both classes of interfaces can exist within the same structure, which would require a more complex thresholding system, though for this work we assume a uniform local atomic environment amongst all grain boundaries. 

For all microstructure characterization tests in this work we employ a $\lambda$ threshold of 0.7, implying that we are defining grain boundaries as local atomic environments showing disorder equivalent to the atomic perturbation observed at around 900K. $\lambda = $ 0.7 was chosen due to the maximum temperatures described in the previous section. As we should expect the grain regions to experience perturbations no greater than those encountered at 600K, those observed at 900K should provide a good approximation for atoms that do not belong to the grains. Once the system has been thresholded, all atoms below the threshold value are removed, leaving only the atoms belonging to the grains. 

This system is then mapped onto a graph, $G$, shown in Fig. \ref{fig:micro-view} (b), where edges are represented by ij pairwise interactions within a 4\AA cutoff radius. A recursive subgraph search algorithm is employed to discover all connected subgraphs, $S_{G}$, within the complete graph $G$. This algorithm is extremely efficient, discovering all subgraphs within a 1.6 million atom system in 1.2 seconds. As all interface atoms were removed prior to the graph construction, all subgraphs in $G$ represent the resulting grains contained within the structure. A 2D slice of the discovered grains in the 3D system is seen in Fig. \ref{fig:micro-view} (b). As each grain is represented by a subgraph in $G$, they can be characterized as graphs. This provides more detail into the grain's shape and connectivity, rather than relying on metrics such as the number of atoms, diameter, etc, which are inherently not unique and do not provide a true quantitative measure of the underlying properties of the structure. Here we use a recently developed graph topology metric \cite{chapman2021sgop} to characterize the subgraphs in $G$. Each edge in the subgraph is labelled according to:

\begin{equation}
	\omega_{i,j} = \frac{1}{d_{i,j}}  \ni d_{i,j} \leq R_{c}
\label{equ:GCN}
\end{equation}

$i$ and $j$ are the atomic indices of the atoms (note that self-interaction terms are not allowed). $d_{i,j}$ is defined as the $l^{2}$-norm between two atoms. $R_{c}$ is the cutoff radius, which was chosen as 4\AA. $\omega_{i,j}$ represents the weight of a given edge for a specific pair of adjacent nodes in the graph. The degree of each node is finally defined as the sum of the elements in a node's edge set, $d_{i} = \sum_{j}^{J} \omega_{i,j}$, where $J$ is the set of all neighbors of atom $i$. The degree sets are then fed into the scalar graph order parameter (SGOP) \cite{chapman2021sgop} scheme for the final characterization of the subgraph. The SGOP functional form is defined as:

\begin{equation}
    \theta_{S_{G},R_{c}} =   \sum_{m}^{D_{s}}P(d_{m})\log_{b}P(d_{m}) + d_{m}P(d_{m})
    \label{eq:sgop}
\end{equation}

where $D_{s}$ is the set of unique node degrees in a subgraph, with $P_{d_{m}}$ being the probability of a given degree, $d_{m}$, occurring in the subgraph. As discussed in previous works, the SGOP value provides a measure of the shape and connectivity of the graph at a structural level. In this way we can observe, quantitatively, the difference between grains with a similar number of atoms, diameter, circumference, density, etc. 


\section*{Supporting Information}
This work contains supplemental information which can be found online.

\section*{Data Availability}
All data required to reproduce this work can be requested by contacting the corresponding author.

\section*{Acknowledgements}
J. Chapman, T. Hsu, X. Chen, T. W. Heo, and B. C. Wood are partially supported by the Laboratory Directed Research and Development (LDRD) program (20-SI-004) at Lawrence Livermore National Laboratory. This work was performed under the auspices of the US Department of Energy by Lawrence Livermore National Laboratory under contract No. DE-AC52-07NA27344.

\section*{Author Contributions}
X. Chen and B. C. Wood supervised the research. J. Chapman performed all MD simulations, and devised/implemented the autonomous microstructure feature extraction methodology. T. Hsu trained the GNN and performed all GNN-related predictions. J. Chapman and T. Hsu devised the theoretical SODAS framework. B. Wood, T. Hsu, and J. Chapman devised the atoms-to-field mapping, while T. Hsu implemented it. T. W. Heo provided insight into the connection between atomistic and phase field modelling, and helped guide discussions surrounding the atoms to continuous field methodology. J. Chapman and T. Hsu wrote the manuscript with inputs from all authors.

\section*{Competing Interests}
The authors declare no competing financial or non-financial interests.
\printbibliography

\end{document}


\maketitle

\clearpage

\renewcommand{\thepage}{S\arabic{page}} 
\renewcommand{\thesection}{S\arabic{section}}  
\renewcommand{\thetable}{S\arabic{table}}  
\renewcommand{\thefigure}{S\arabic{figure}}
\setcounter{figure}{0}


\subsection*{Model training}
We used PyTorch Geometric \cite{fey2019fast} to develop the GNN, with the model parameters described in Table~\ref{tab:model-params}. The model was trained with the Adam optimizer \cite{kingma2014adam}, carried out using PyTorch \cite{paszke2019pytorch} and PyTorch Geometric \cite{fey2019fast} on a NVIDIA V100 (Volta) GPU. The binary cross entropy was used as the loss function during training. The training parameters are described in Table~\ref{tab:train-params}. All other parameters, if unspecified in this work, default to values per PyTorch 1.8.1 and PyTorch Geometric 1.7.2.

\begin{figure}
    \centering
    \includegraphics[width=0.5\textwidth]{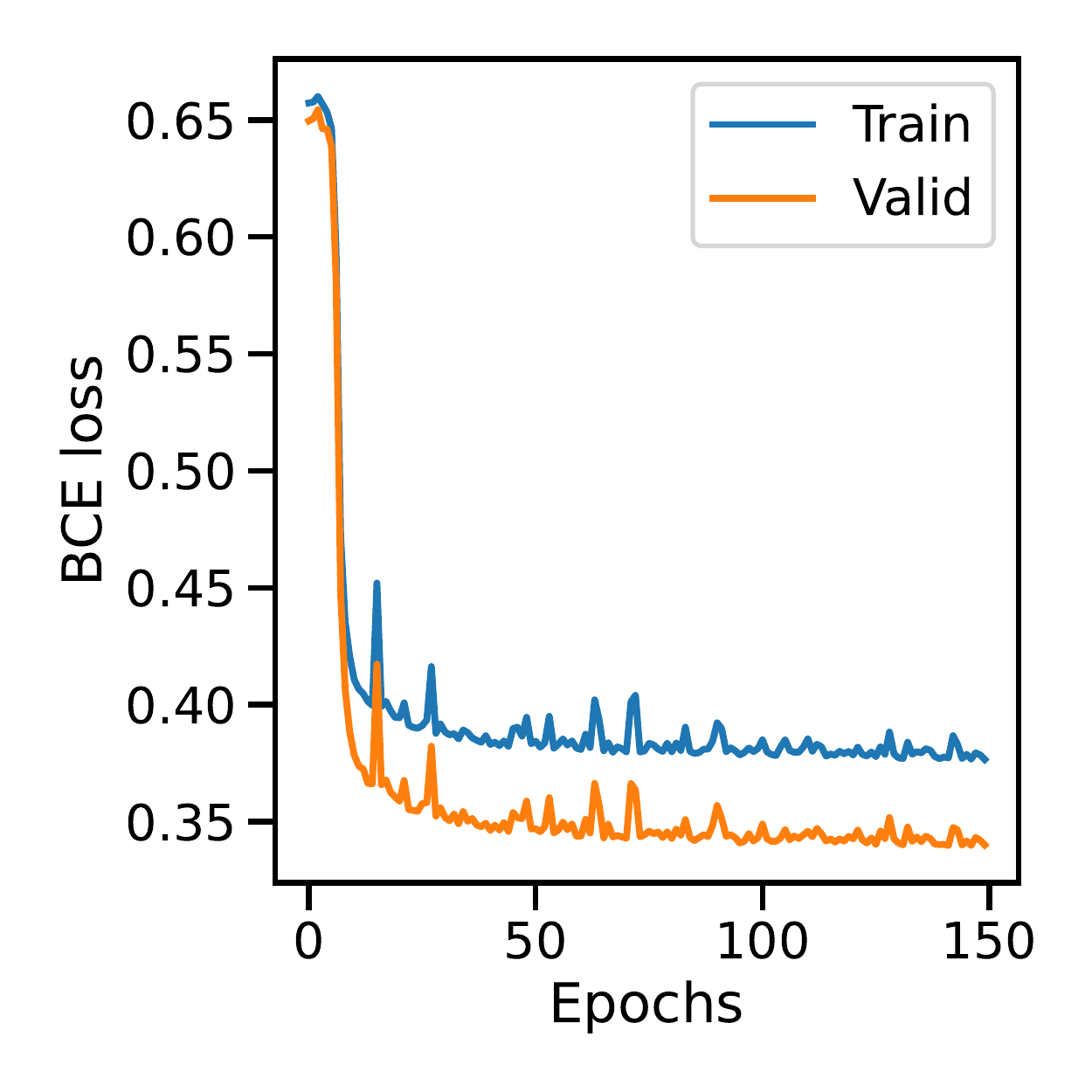}
    \caption{Loss curves during training.}
    \label{fig:loss-curves}
\end{figure}

The training dataset is based on a CMD trajectory of 1000 snapshots, from which 50 were randomly sampled for model validation. The training and validation losses are shown in Fig.~\ref{fig:loss-curves}.

\begin{table}
    \centering
    \caption{Model parameters}
    \begin{tabular}{l l l}
    \hline
    Name & Notation & Value \\
    \hline
    Number of interaction layers            & $L$               & 3                     \\
    RBF cutoff                              & $R_c$             & 3.5 \si{\angstrom}    \\
    Number of channels                      & $D$               & 100                   \\
    \hline
    \end{tabular}
    \label{tab:model-params}
\end{table}

\begin{table}
    \centering
    \caption{Training parameters}
    \begin{tabular}{l l l}
    \hline
    Name & Notation & Value \\
    \hline
    Batch size                          & $M$                       & 16           \\
    Number of epochs                    & $N_\mathrm{ep}$           & 150           \\
    Learning rate                       & $\eta$                    & 0.0001        \\
    First moment coefficient for Adam   & $\beta_1$                 & 0.9           \\
    Second moment coefficient for Adam  & $\beta_2$                 & 0.999         \\
    \hline
    \end{tabular}
    \label{tab:train-params}
\end{table}

\newpage
\subsection*{Model validation}
\begin{figure}
    \centering
    \includegraphics[trim={0 0cm 12cm 0},width=0.5\textwidth]{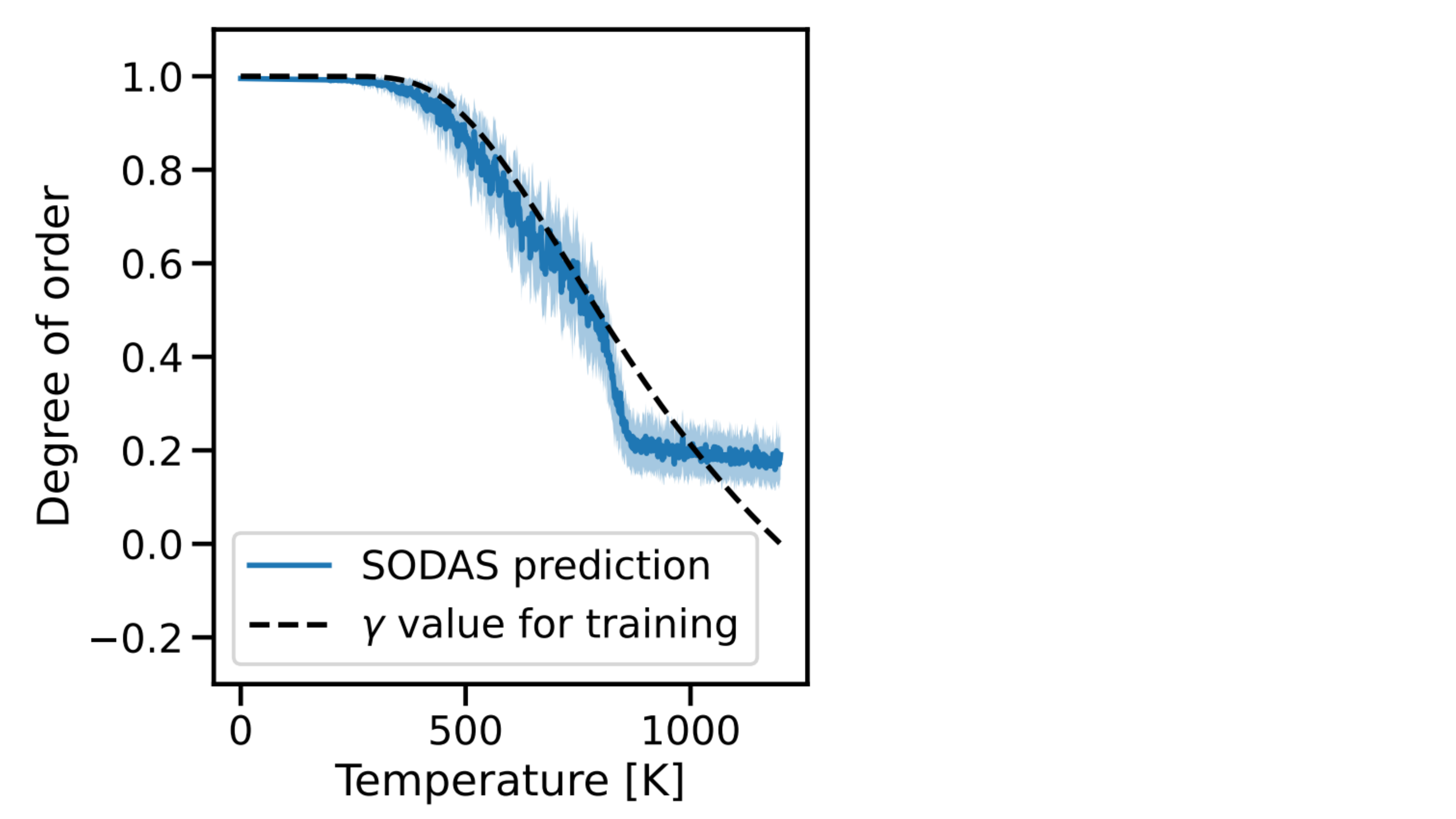}
    \caption{SODAS calculations on bulk structures taken during a superheating MD simulation. Values along the y-axis represent the average SODAS value for each structure in the model's training set. Light blue shaded regions indicate the spread of the atomic SODAS value for a given configuration. The dashed line indicates the theoretical values of $\gamma$ while the plotted SODAS values represent the accuracy of the GNN mapping.}
    \label{fig:gamma_train}
\end{figure}

Fig.~\ref{fig:gamma_train} compares the GNN learned SODAS values on the training data with respect to the theoretical values for $\gamma$. Here, we can see excellent agreement between the GNN mapping and $\gamma$ up to around $T = $700K. This indicates that our MD simulations are sufficient to capture the configurational entropy present within the material at these temperatures. Deviations begin around 700K, with a sharper decrease in the predicted average SODAS value between 700K and 900K. This change in the slope could indicate that there exist significantly more disorder present within the MD-generated configurations within this temeparture range than exist in reality. As such, this deviation can be attributed to a sampling problem, with a more robust representation of structures between 700K and 900K being required to more accurately learn $\gamma$.

There are also deviations above $T = $1000K, with a nearly identical average SODAS value predicted between 900K and 1200K. This can be explained by our choice of $T_m =$ 1200K, and the true melting temperature of the chosen EAM potential (~1050K). Therefore, above 1050K there exists a nearly identical level of local disorder compared to structures at 1200K. Interestingly, the GNN SODAS mapping inherently learns this, providing a potential feedback mechanism by which to tune the value of $T_m$ for a given material system.

\printbibliography